\documentclass[11pt,a4paper,liststotoc,idxtotoc]{scrartcl} 

\usepackage[english]{babel} 
\usepackage[utf8]{inputenc} 
\usepackage[T1]{fontenc}
\usepackage{amsmath,amssymb,amsfonts} 
\usepackage{graphicx} 
\addtokomafont{sectioning}{\rmfamily} 
\usepackage{booktabs, tabularx} 
\usepackage[left=2cm, right=2cm, top=2cm, bottom=2cm]{geometry}
\usepackage[onehalfspacing]{setspace} 
\usepackage{acronym}
\usepackage{float}
\usepackage{siunitx}
\usepackage{enumerate}
\usepackage{hyperref} 
\usepackage{harvard} 
\usepackage{algorithm}
\usepackage{geometry}
\usepackage{pdflscape}
\usepackage[noend]{algpseudocode}

\newcommand{\citeN}[1]{\citeasnoun{#1}}

\usepackage{authblk}
\title{\Large{Fast and Feasible Estimation of Generalized Linear Models with High-Dimensional $k$-way Fixed Effects\footnote{An earlier version of the paper is named \textit{Fast and Feasible Estimation of Generalized Linear Models with Many Two-Way Fixed Effects}  \href{https://arxiv.org/pdf/1707.01815.pdf}{\url{https://arxiv.org/pdf/1707.01815.pdf}}   }}}
\author[]{\large{Amrei Stammann}\footnote{Email: Amrei.Stammann@hhu.de} }
\affil[]{\large{Heinrich-Heine University Düsseldorf}}
\date{\today}

\usepackage{chngcntr} 
\counterwithout{footnote}{section}

\numberwithin{equation}{section} 
\numberwithin{table}{section}

\begin{document}

\maketitle	
\thispagestyle{empty} 

\begin{abstract}
We present a fast and memory efficient  algorithm for the  estimation of generalized linear models with an additive separable $k$-way error component. The brute force approach uses dummy variables to account for the unobserved heterogeneity, but quickly faces computational limits. Thus, we show how a weighted version of the Frisch-Waugh-Lovell theorem combined with the method of alternating projections can be incorporated into a Newton-Raphson algorithm to dramatically reduce the computational costs.  The algorithm is  especially useful in situations, where generalized linear models with $k$-way fixed effects  based on dummy variables are computationally demanding or even infeasible due to time or memory limitations. In a simulation study and an empirical application we demonstrate the performance of our algorithm. 
  \\
  \vfill{}
\end{abstract}

\newpage

\section{Introduction}

Fixed effects models are popular specifications to account for unobserved heterogeneity; for example in labor economics often worker, firm and/or time fixed effects are used, and in trade economics  importer-time, exporter-time, and dyadic fixed effects are required to estimate structural gravity models.
Especially in large micro-level panels like  the U.S. PSID  or pseudo-panels of trade flows like the CEPII such model specifications can lead to high-dimensional fixed effects. 

Usually the unobserved heterogeneity is captured by including a dummy variable for each level of each fixed effects category. In classical one-way linear regression models it is possible to use a computational trick known as demeaning or with-in transformation to get rid of these dummy variables. 
Even if only one fixed effects dimension is large it is straightforward to add the smaller dimensions as dummy variables to the regressor matrix and to demean over the larger fixed effects dimension.
If all or many fixed effects dimensions are large the aforementioned approach would require the generation and inversion of a potentially large regressor matrix.  To tackle this computational burden various algorithms have been proposed  (see among others \citeN{Guimaraes2010}, \citeN{Gaure2013theory}, \citeN{Correia2016}, \citeN{Somaini2016}). These algorithms  rely on the Frisch-Waugh-Lovell (FWL) theorem (\citeN{Frisch1933}, \citeN{Lovell1963}) and have been so far  developed in particular for linear models.

In the case of generalized linear models, for instance probit and logit models, no general efficient $k$-way fixed effects algorithms have been designed yet. Like in the linear case it is possible to include the lower fixed effects dimensions as additional regressors and to apply special algorithms to concentrate out the larger dimension. For example the partitioned inverse approach proposed by \citeN{Chamberlain1980}  or the algorithm proposed by \citeN{Stammann2016} are suited.  \citeN{Guimaraes2010} suggest a Gauss-Seidel algorithm to estimate linear and non-linear models with high-dimensional  fixed effects.\footnote{For linear models they also sketch an alternative efficient algorithm based on the method of alternating projections. However,  it is not declared as  an alternating projection approach.   \citeN{Gaure2013theory}  was the first one who introduced the method of alternating projections in the context of linear regression model with high-dimensional fixed effects along with an extensive theoretical foundation.} Using the example of poisson  regression  they show how a closed form of the fixed effects can be abused to derive an efficient algorithm. However, most generalized linear models do not have such a closed form. For these cases \citeN{Guimaraes2010} show that  the Gauss-Seidel algorithm can be combined with a demanding numerical optimization routine to solve for the fixed effects.\footnote{The author of this paper has been made aware of a Stata routine \textit{poi2hdfe} written by Paulo Guimaraes to estimate two-way fixed effects poisson models.    This routine is not based on the Gauss-Seidel algorithm mentioned previously, but rather uses  the method of alternating projections by incorporating the Stata routine \textit{hdfe} of \citeN{Correia2016} into an iteratively reweighted least squares  algorithm. To the best of our knowledge  the implemented routine has not been presented in a paper yet. The underlying approach is similar, albeit different, to the one we present.}
Recently, \citeN{Larch2017} modified the Gauss-Seidel algorithm   of \citeN{Guimaraes2010} for poisson models to estimate a gravity model with a high-dimensional three-way fixed effects specification.

We derive a straightforward and memory efficient  maximum likelihood  approach that  can be applied to all generalized linear models with a $k$-way error component.\footnote{A first version of our algorithm is available as an R-package \textit{alpaca} (co-authored with Daniel Czarnowske) which can be downloaded here: \href{https://github.com/amrei-stammann/alpaca}{\url{https://github.com/amrei-stammann/alpaca}}. Note that \textit{alpaca} only provides routines for non-linear GLM's because there is already a comprehensive R-package \textit{lfe} by Simen Gaure for linear regression models \cite{Gaure2013lfe}. } 
Our starting point is an algorithm proposed by \citeN{Gaure2013theory} for linear regression models with high-dimensional fixed effects which uses an iterative demeaning procedure based on alternating projections. We extend it to  generalized linear models by using a result previously shown by \citeN{Stammann2016} which allows the application of the FWL theorem in each iteration of the Newton-Raphson optimization routine. This results in an efficient approach where the fixed effects are concentrated out of the parameters update. We refer to this step as pseudo-demeaning. Unlike in linear models, where the structural parameters can be estimated separably from the fixed effects,  this does not hold for for generalized linear models. Since  the fixed effects contribute to the linear predictor they have to be updated in each iteration of the optimization routine.  Fortunately, it turns out to be  much less computational challenging to update  the linear predictor itself.  Nevertheless, we show how  the estimates of the fixed effects can be recovered efficiently ex-post.  

The standard approach to estimate generalized linear models is an iteratively reweighted least squares (IRLS) algorithm.
In contrast to  the state-of-the-art, our routine is based on the classical Newton-Raphson formulation.\footnote{Note that IRLS is derived from Newton-Raphson by reformulating the Newton step as a weighted least squares step with an adjusted response.} This has the advantage that the scores of the log-likelihood can be  obtained directly from the optimization procedure without  additional post-estimation procedures. The scores are required to  compute robust and (multi-way) clustered standard errors. 


The remainder of the paper is organized as follows. First we introduce the $k$-way fixed effects generalized linear model, and show that the Newton-Raphson update is just a weighted least squared problem where we can apply the FWL theorem. Next we combine the resulting projection matrix with the method of alternating projections to arrive at a straightforward  pseudo-demeaning algorithm that will be incorporated into a standard Newton-Raphson routine. Afterwards we present two efficient ways to recover the fixed effects ex-post. Finally, a simulation study highlights the performance of our Newton-Raphson pseudo-demeaning algorithm and an empirical example in trade economics demonstrates a possible area of application.

\section{The Model}
A generalized linear model consists of three parts: a stochastic component $\boldsymbol{\mu}$, a systematic component $\boldsymbol{\eta}$, and a link $h(\cdot)$ between both components \cite{Mccullagh1989}. 

In a $k$-way fixed effects generalized linear model the linear predictor takes the following specific form:
\begin{equation}
	\label{eq:eta}
	\boldsymbol{\eta}= \mathbf{Z} \boldsymbol{\gamma} =  \mathbf{D} \boldsymbol{\alpha}     + \mathbf{X} \boldsymbol{\beta} = \sum_{k=1}^K \mathbf{D}_k \boldsymbol{\alpha}_k     + \mathbf{X} \boldsymbol{\beta} \, ,
\end{equation}
where the regressor matrix $\mathbf{Z}$ can be split into a sparse part $\mathbf{D}$ and a remaining part  $\mathbf{X}$. More specifically, the matrices $\mathbf{D}_k$ arise from dummy encoding $K$ categorical variables and capture the unobserved heterogeneity. Each dummy matrix is of dimension $(n \times l_{k})$, where $n$ is the number of observations and $l_{k}$ is the number of levels of the $k$-th categorical variable. The corresponding parameters $\boldsymbol{\alpha} = [\boldsymbol{\alpha}_{1}, \ldots, \boldsymbol{\alpha}_{K}]^{\prime}$ are called fixed effects. The remaining part $\mathbf{X}$ is a $(n \times p)$ matrix of variables of interest and the corresponding parameters $\boldsymbol{\beta}$ are the structural parameters.

The further components of the model can be expressed as follows:
\begin{equation}
\mathbf{E(y)}=\boldsymbol{\mu}=h^{-1}(\boldsymbol{\eta}) \nonumber \, ,
\end{equation}
where the link function $h(\cdot)$ is a monotonic differentiable function and $\mathbf{y}$ is a realization of an independently distributed random variable from the exponential family $\mathbf{Y}$. The distribution is given by:
\begin{equation}
	f_{Y}(y, \theta, \phi) = \exp\left((y \theta - b(\theta)) / a(\phi) + c(y, \phi)\right) \, ,
\end{equation}
where $a(\cdot)$, $b(\cdot)$, and $c(\cdot)$ are specific functions. We consider the cases where $\phi$ is known and thus $\theta$ is a canonical parameter.
Table 2.1 summarizes the corresponding functions and parameters used in this paper (logit and poisson). For other generalized linear models please consult e.g. \citeN{Mccullagh1989}.
\begin{table}[H]
	\caption{Logit and Poisson}
	\begin{center}
		\begin{tabular}{lcc}
			\toprule
			&          \text{Logit}            & \text{Poisson} \\ \midrule
			\text{Dispersion parameter} $\phi$   &                1                      &       1        \\
			\text{Cumulant function} $b(\theta)$ &     $\log(1+\exp(\theta))$          & $\exp(\theta)$ \\
			$c(y, \phi)$                         &               $0$                    &      $\log(y!)$      \\
			$\mu(\theta)$                        & $\exp(\theta)/(1+\exp(\theta))$        & $\exp(\theta)$ \\
			\text{Canonical link} $\theta(\mu)$  &       $\log(\mu/(1-\mu))$             &  $\log(\mu)$   \\
			\text{Variance function} $V(\mu)$    &          $\mu (1-\mu)$             &     $\mu$      \\ \bottomrule
		\end{tabular}
	\end{center}
	\footnotesize{\emph{Note}: Following \citeN{Mccullagh1989} table 2.1.}
\end{table}
The unknown parameters $\boldsymbol{\gamma}=[\boldsymbol{\alpha}, \boldsymbol{\beta}]'$ are estimated using maximum likelihood. The log-likelihood is 
\begin{equation}
	\mathcal{L} = \sum_{i = 1}^{n}(y_{i} \theta_{i} - b(\theta_{i})) / a(\phi) + c(y_{i}, \phi) \, ,
\end{equation}
which can be maximized iteratively. The Newton-Raphson update is
\begin{align}
\label{eq:dummynewton}
	\boldsymbol{\gamma}^{r}-\boldsymbol{\gamma}^{r-1}=-(\mathbf{H}^{r-1})^{-1}\mathbf{g}^{r-1} \, ,
\end{align}
where $\mathbf{g}^{r}$ and $\mathbf{H}^{r}$ are the gradient and Hessian at iteration $r$.

Since $\theta(\boldsymbol{\mu})$ is the canonical link we can apply the chain rule which leads to the following expression of the gradient:
\begin{equation}
\label{eq:hessian}
	\frac{\partial \mathcal{L}}{\partial \boldsymbol{\gamma}^{r}} = \mathbf{g}^{r} = \mathbf{Z}^{\prime} \mathbf{W}^{r} \boldsymbol{\nu}^r \, ,
\end{equation}
where $ \boldsymbol{\nu}^r =  \left( (\mathbf{y} - \boldsymbol{\mu}^{r}) \odot \frac{\partial \boldsymbol{\eta}^{r}}{\partial \boldsymbol{\mu}^{r}}\right)$, $\mathbf{W}^{r}$ is a positive definite diagonal weighting matrix with its $i$-th entry equal to $\left(\frac{\partial \mu_{i}^{r}}{\partial \eta_{i}^{r}}\right)^2 / V_{i}^{r} = 1 / (\left(\frac{\partial \eta_{i}^{r}}{\partial \mu_{i}^{r}}\right)^{2} V_{i}^{r})$. The Hessian can be derived in the same manner:
\begin{equation}
\label{eq:fullhessian}
	\frac{\partial^{2} \mathcal{L}}{\partial \boldsymbol{\gamma}^{r} \partial \boldsymbol{\gamma}^{r \prime}} = \mathbf{H}^{r} = - \mathbf{Z}^{\prime} \mathbf{W}^{r} \mathbf{Z} \, .
\end{equation}
For now we assume $\mathbf{Z}$ to have full rank and $dim(\mathbf{Z})= n \times(p + l)$, where $l \leq \sum_{k=1}^{K}l_k$ denotes the columns of the sparse part of $\mathbf{Z}$. Later this assumption will be relaxed. \footnote{Usually the sparse part of $\mathbf{Z}$ has no full rank, such that some columns are removed for the estimation. For example in the classical two-way fixed effects model with individual and time fixed effects, one column of the dummy matrix has to be removed.}

Brute-force implementation of (\ref{eq:dummynewton}) would require the computation and inversion of a potentially large Hessian of dimension $(p+l) \times (p+l) $ which quickly becomes computationally demanding or even infeasible.

In the next section we present a new Newton-Raphson pseudo-demeaning algorithm based on the Frisch-Waugh-Lovell (FWL) theorem in combination with the method of alternating projections. This approach substantially decreases the computational costs of the brute-force implementation.

\section{The Pseudo-Demeaning Algorithm}
\label{sec:pseudodemeaning}
\subsection{The FWL Theorem and the Newton-Raphson Update}
In the classical fixed effects linear model the FWL theorem is applied to separate the estimation of the fixed effects from the structural parameters.
Recently \citeN{Stammann2016} showed how the FWL theorem can be adapted to separate the Newton-Raphson updates of the structural parameters from the fixed effects updates in a one-way fixed effects logit model. The same logic can be applied to $k$-way fixed effects generalized linear models. The parameter update \label{eq:newton} is essentially the solution of  a weighted least squares problem:\footnote{ The standard IRLS reformulation would be $
	\boldsymbol{\gamma}^{r}=(\mathbf{Z}^{\prime} \mathbf{W}^{r - 1} \mathbf{Z})^{-1}\mathbf{Z}^{\prime} \mathbf{W}^{r - 1} \left( \boldsymbol{\nu}^{r-1} + \mathbf{Z} \boldsymbol{\gamma}^{r-1} \right)$. We use the different formulation (\ref{eq:startingpoint}) in order to obtain the scores of the log-likelihood directly from the estimation routine.}
\begin{align}
\label{eq:startingpoint}
\boldsymbol{\gamma}^{r}-\boldsymbol{\gamma}^{r-1}&=(\mathbf{Z}^{\prime} \mathbf{W}^{r - 1} \mathbf{Z})^{-1}\mathbf{Z}^{\prime} \mathbf{W}^{r - 1} \boldsymbol{\nu}^{r-1} \, .
\end{align}
Thus the parameter update can be obtained by the following regression
\begin{equation}
\label{eq:regression}
\tilde{\boldsymbol{\nu}}^{r - 1}= \widetilde{\mathbf{D}}^{r - 1} (\boldsymbol{\alpha}^{r} - \boldsymbol{\alpha}^{r - 1}) + \widetilde{\mathbf{X}}^{r - 1} (\boldsymbol{\beta}^{r}-\boldsymbol{\beta}^{r - 1}) \, , 
\end{equation}
where  $\tilde{\boldsymbol{\nu}}^{r} = \widetilde{\mathbf{W}}^{r} \left( (\mathbf{y} - \boldsymbol{\mu}^{r}) \odot \frac{\partial \boldsymbol{\eta}^{r}}{\partial \boldsymbol{\mu}^{r}}\right)$, $\widetilde{\mathbf{D}}^{r} = \widetilde{\mathbf{W}}^{r} \mathbf{D}$,  $\widetilde{\mathbf{X}}^{r} = \widetilde{\mathbf{W}}^{r} \mathbf{X}$, and $\widetilde{\mathbf{W}}^{r} = (\mathbf{W}^{r})^{1 / 2}$.

This transformation of the update formula  allows to eliminate $\widetilde{\mathbf{D}}^{r - 1} (\boldsymbol{\alpha}^{r}-\boldsymbol{\alpha}^{r - 1})$ from (\ref{eq:regression}) via the FWL theorem:
\begin{align}
\label{eq:projectedsystem}
\mathbf{M}_{\widetilde{\mathbf{D}}}^{r - 1} \tilde{\boldsymbol{\nu}}^{r - 1} &= \mathbf{M}_{\widetilde{\mathbf{D}}}^{r - 1} \widetilde{\mathbf{D}}^{r - 1} (\boldsymbol{\alpha}^{r} - \boldsymbol{\alpha}^{r - 1}) + \mathbf{M}_{\widetilde{\mathbf{D}}}^{r - 1} \widetilde{\mathbf{X}}^{r - 1} (\boldsymbol{\beta}^r - \boldsymbol{\beta}^{r - 1}) \\ 
&= \mathbf{M}_{\widetilde{\mathbf{D}}}^{r - 1} \widetilde{\mathbf{X}}^{r - 1} (\boldsymbol{\beta}^r - \boldsymbol{\beta}^{r - 1}) \, , \nonumber
\end{align}
where the annihilator matrix $\mathbf{M}_{\widetilde{\mathbf{D}}}^{r} = \mathbf{I}_{n} - \widetilde{\mathbf{D}}^{r}(\widetilde{\mathbf{D}}^{r \prime} \widetilde{\mathbf{D}}^{r})^{-1} \widetilde{\mathbf{D}}^{r \prime}$ is the projection onto the orthogonal complement of the column space of $\widetilde{\mathbf{D}}^{r}$.\footnote{\label{foot:wlsprojection1} Note, $\mathbf{M}$ is idempotent and that   (\ref{eq:projectedsystem}) can be transformed into
$\widetilde{\mathbf{W}}^{r - 1}\mathbf{P}^{r - 1} \boldsymbol{\nu}^{r - 1} = \widetilde{\mathbf{W}}^{r - 1}\mathbf{P}^{r - 1}  \mathbf{X}^{r - 1} (\boldsymbol{\beta}^r - \boldsymbol{\beta}^{r - 1})$,
where  $\mathbf{P}^{r} = \mathbf{I}_n - \mathbf{D}(\mathbf{D}^{\prime} \mathbf{W}^{r} \mathbf{D})^{-1} \mathbf{D}^{\prime} \mathbf{W}^{r}$.
Both projection approaches  are suitable to concentrate out the high-dimensional fixed effects from (\ref{eq:regression}). For the rest of the paper we restrict ourselves to the first one (see footnote \ref{foot:wlsprojection2}).}

We call $\mathbf{M}_{\widetilde{\mathbf{D}}}^{r - 1} \tilde{\boldsymbol{\nu}}^{r - 1}$ and  $\mathbf{M}_{\widetilde{\mathbf{D}}}^{r - 1} \widetilde{\mathbf{X}}^{r -1}$ the pseudo-demeaned variables that can be used to compute the update of the structural parameters separately from the high-dimensional fixed effects updates at very low computational costs. However, the brute-force pseudo-demeaning ends up in a computational challenge itself since the annihilator matrix $\mathbf{M}_{\widetilde{\mathbf{D}}}^{r}$ has dimension $(n \times n)$ and is typically non-sparse. One exception is the case  $K = 1$ where the block-diagonal structure of $(\widetilde{\mathbf{D}}^{r \prime} \widetilde{\mathbf{D}}^{r})^{-1}$ allows to derive a straightforward formula to compute the pseudo-demeaned variables without costly matrix operations \cite{Stammann2016}. For $K > 1$ this is not possible since $(\widetilde{\mathbf{D}}^{r \prime} \widetilde{\mathbf{D}}^{r})^{-1}$ looses its sparse structure. Fortunately, we can use a combination of the one-way pseudo-demeaning along with the method of alternating projections to approximate the pseudo-demeaned variables directly without having to compute the expensive annihilator matrix $\mathbf{M}_{\widetilde{\mathbf{D}}}^{r}$.

\subsection{The Method of Alternating Projections}
An approach to compute the pseudo-demeaned variables efficiently is a method called alternating projections (AP) tracing back to \citeN{Neumann1950} and \citeN{Halperin1962}. \citeN{Gaure2013theory} introduced AP in the context of classical linear models with many fixed effects categories. We show how the AP approach can be adapted to generalized linear models.  

In order to introduce the alternating projection methods we first have to consider some basics from linear algebra. Let $R(\cdot)$ denote the column space and $R(\cdot)^{\perp}$ denotes its orthogonal complement.
Suppose we want to compute $\ddot{\mathbf{v}}=\mathbf{M}_{\widetilde{\mathbf{D}}} \mathbf{v}$ where $\mathbf{v}$ is an arbitrary $(n \times 1)$ vector. Since $\mathbf{M}_{\widetilde{\mathbf{D}}}$ is the projection onto the orthogonal complement of the column space of $\widetilde{\mathbf{D}}$, $\ddot{\mathbf{v}} \in R(\widetilde{\mathbf{D}})^{\perp}$. The column space of $\widetilde{\mathbf{D}}$ is the intersection of the column spaces of the weighted dummy matrices $\widetilde{\mathbf{D}}_{k}$, i.e. $R(\widetilde{\mathbf{D}}) = \cap_{k=1}^K R(\widetilde{\mathbf{D}}_{k})$. The same is true for the orthogonal complement $R(\widetilde{\mathbf{D}})^{\perp} = \cap_{k=1}^K R(\widetilde{\mathbf{D}}_{k})^{\perp}$. Altogether, the pseudo-demeaned variable lies in the intersection of the subspaces $R(\widetilde{\mathbf{D}}_{k})^{\perp}$,  $\ddot{\mathbf{v}} \in \cap_{k=1}^K R(\widetilde{\mathbf{D}}_{k})^{\perp}$. Since alternating projection (AP) methods  are used to approximate a point in the intersection of a finite number of closed subspaces of a Hilbert space (see \citeN{Escalante2011alternating}) they are suitable to find $\ddot{\mathbf{v}}$. The idea is to approximate $\mathbf{M}_{\widetilde{\mathbf{D}}} \mathbf{v}$ by projecting repeatedly  on the individual subspaces $R(\widetilde{\mathbf{D}}_{k})^{\perp}$ which are, in general, much easier to compute.

There are basically two AP methods which differ by how the individual projections are linked: Neumann-Halperin and Cimmino. \citeN{Neumann1950} developed the AP method for the case of two subspaces, and \citeN{Halperin1962} extended this to a finite number of subspaces. Originally the method proposed by \citeN{Cimmino1938}  is intended to solve linear systems of equations. However, as shown by \citeN{Kammerer1972} it is also suitable for linear operations on subspaces (see \citeN{Hernandez2011}).

The Neumann-Halperin approach can be summarized as follows
\begin{equation}
  \lim_{N \rightarrow \infty} \lVert (\mathbf{M}^{r}_{\widetilde{\mathbf{D}}_{1}^{r}}  \mathbf{M}^{r}_{\widetilde{\mathbf{D}}_{2}^{r}}\cdots \mathbf{M}^{r}_{\widetilde{\mathbf{D}}_{K}^{r}})^{N} \mathbf{v} - \mathbf{M}_{\widetilde{\mathbf{D}}}^{r} \mathbf{v} \rVert = 0 \, . \nonumber
\end{equation}
This means, that  $\mathbf{v}$ is  projected onto $R(\widetilde{\mathbf{D}}_{1})^{\perp}$, giving some vector $\mathbf{v}_1 \in R(\widetilde{\mathbf{D}}_{1})^{\perp} $. $\mathbf{v}_1 $ is projected  onto $R(\widetilde{\mathbf{D}}_{2})^{\perp}$, giving some vector $\mathbf{v}_2 \in R(\widetilde{\mathbf{D}}_{2})^{\perp} $, which is projected onto the next subspace, and so on, until we project from $R(\widetilde{\mathbf{D}}_{K-1})^{\perp}$ onto $R(\widetilde{\mathbf{D}}_{K})^{\perp}$. This procedure is repeated until convergence.

In contrast to Neumann-Halperin's approach, Cimmino's projections are not nested. Instead one projects $\mathbf{v}$ separately onto each of the $K$ subspaces $R(\widetilde{\mathbf{D}}_{k})^{\perp}$  and computes the centroid of these projections according to: 
\begin{equation}
\lim_{N \rightarrow \infty} \lVert ( \frac{1}{K} \sum_{k=1}^{K} \mathbf{M}^{r}_{\widetilde{\mathbf{D}}_{k}^{r}}  )^{N} \mathbf{v} - \mathbf{M}_{\widetilde{\mathbf{D}}}^{r} \mathbf{v} \rVert = 0 \, . \nonumber
\end{equation}

With help of AP methods the large and non-sparse projection $\mathbf{M}_{\widetilde{\mathbf{D}}}^{r} \mathbf{v}$ can be decomposed into an iterative procedure based on only sparse projections $\mathbf{M}^{r}_{\widetilde{\mathbf{D}}_{k}^{r}} = \mathbf{I}_n - \widetilde{\mathbf{D}}_{k}^{r} (\widetilde{\mathbf{D}}_{k}^{r \prime} \widetilde{\mathbf{D}}_{k}^{r})^{- 1} \widetilde{\mathbf{D}}_{k}^{r \prime}$ which  translate into  one-way pseudo-demeaning over category $k$. Using the result shown by \citeN{Stammann2016}, the projections $\mathbf{M}^{r}_{\widetilde{\mathbf{D}}_{k}^{r}} \mathbf{v}$ can be efficiently computed as follows:\footnote{\label{foot:wlsprojection2}It would also be possible to use the alternative projection defined in footnote \ref{foot:wlsprojection1}. Although this projection seems to be favorable due to fewer operations, we found that it often takes longer to pseudo-demean such that in total none of the projections is superior with respect to total computation time.}
\begin{equation}
\label{eq:pseudodemeaning}
(\mathbf{M}^{r}_{\widetilde{\mathbf{D}}_{k}^{r}} \mathbf{v})_{i} = v_i -  \tilde{w}_i^{r}  \frac{\sum_{j \in g_{k \kappa}} \tilde{w}_j^r v_j}{\sum_{j \in g_{k \kappa}} w_j^r} \quad \forall i \in g_{k\kappa}\, ,
\end{equation}
where $g_{k\kappa}$ defines a group consisting of those observations that share the same level $\kappa$ in category $k$, and $\tilde{w}_i^{r}$ and $w_i^{r}$ are the $i$-th diagonal entry of  $\widetilde{\mathbf{W}}^{r}$ and $\mathbf{W}^{r}$ respectively. Equation (\ref{eq:pseudodemeaning}) demonstrates that the individual projections  essentially subtract ``weighted'' group means from the dependent variable $\tilde{\boldsymbol{\nu}}^{r}$ and the regressor matrix $\widetilde{\mathbf{X}}^{r}$.

In order to approximate $\mathbf{M}_{\widetilde{\mathbf{D}}}^{r}\tilde{\boldsymbol{\nu}}^{r}$ and $\mathbf{M}_{\widetilde{\mathbf{D}}}^{r} \widetilde{\mathbf{X}}^{r}$, the alternating projection algorithm  is subsequently applied to $\tilde{\boldsymbol{\nu}}$ and each column of $\widetilde{\mathbf{X}}$. This could be either the Neumann-Halperin algorithm (Algorithm 1) or the Cimmino algorithm (Algorithm 2). 

\begin{algorithm}
	\caption{Pseudo-Demeaning: Neumann-Halperin}
	\begin{algorithmic}[1]
		\State Let $\mathbf{v} \in \{\tilde{\boldsymbol{\nu}}^{r}, \tilde{\mathbf{x}}_j^{r}\}$, $j=1, \dots, p$.
		\State Set $i=1$ and $\mathbf{z}_i= \mathbf{v}$.
		\Repeat 
	\State Set $\mathbf{z}_{i0}=\mathbf{z}_i$.
	\For{$k=1, \dots, K$}
	\State Compute $\mathbf{z}_{ik}$ by subtracting the ``weighted'' group mean from $\mathbf{z}_{i(k-1)}$  (see formula \ref{eq:pseudodemeaning}).
	\EndFor	
	\State Set $i=i+1$, $\mathbf{z}_{i}=\mathbf{z}_{iK}$.
	\Until{\textbf{convergence}.}
	\State Set $\ddot{\mathbf{v}}=\mathbf{z}_i $.
	\end{algorithmic}
\end{algorithm}


\begin{algorithm}
	\caption{Pseudo-Demeaning: Cimmino}
	\begin{algorithmic}[1]
		\State Let $\mathbf{v} \in \{\tilde{\boldsymbol{\nu}}^{r}, \tilde{\mathbf{x}}_j^{r}\}$, $j=1, \dots, p$.
		\State Set $i=1$, $\mathbf{z}_i= \mathbf{v}$, and $\mathbf{z}_{sum} = \mathbf{0}_{p}$.
		\Repeat
		\State Set $\mathbf{z}_{i0}=\mathbf{z}_i$.
		\For{$k=1, \dots, K$}
		\State Compute $\mathbf{z}_{ik}$ by subtracting the ``weighted'' group mean from $\mathbf{z}_{i0}$  (see formula \ref{eq:pseudodemeaning}).
		\State $\mathbf{z}_{sum} = \mathbf{z}_{ik} + \mathbf{z}_{sum} $
		\EndFor	
		\State Set $i=i+1$, $\mathbf{z}_{i}=\frac{1}{K} \mathbf{z}_{sum}$.
		\Until{\textbf{convergence}.}
		\State Set $\ddot{\mathbf{v}}=\mathbf{z}_i $.
	\end{algorithmic}
\end{algorithm}

Afterwards, the approximations $\ddot{\boldsymbol{\nu}}^{r}$ and $\ddot{\mathbf{X}}^{r}$ are  used to compute the updates of the structural parameters efficiently:
\begin{equation}
\label{eq:betaupdate}
(\boldsymbol{\beta}^r-\boldsymbol{\beta}^{r-1})=(\ddot{\mathbf{X}}^{r - 1 \prime}\ddot{\mathbf{X}}^{r - 1})^{-1}\ddot{\mathbf{X}}^{r - 1\prime}\ddot{\boldsymbol{\nu}}^{r - 1} \, .
\end{equation}

Note that we do not require  the full rank  assumption of $\mathbf{D}$ and $\widetilde{\mathbf{D}}$ anymore. Let $\widetilde{\mathbf{\mathcal{D}}}$ denote the rank deficient weighted dummy matrix where no collinear columns have been removed. The structural parameter updates (\ref{eq:betaupdate}) are not influenced by the design of the dummy variable matrix, since $R(\widetilde{\mathbf{D}})^{\perp} = R(\widetilde{\mathbf{\mathcal{D}}})^{\perp}$ and thus  $\tilde{\boldsymbol{\nu}}$ and $\widetilde{\mathbf{X}}$ are projected onto the correct space anyway.  For simplicity we do not further distinguish whether $\mathbf{D}$ and $\widetilde{\mathbf{D}}$ are rank deficient or not.\footnote{What is still required is that $\mathbf{X}$ has full rank and that none of the regressors is perfectly collinear with the fixed effects.  Whereas the former is easy to check the latter implies the need of a well-thought-out model specification by the researcher. In these cases it is unlikely that the estimation routine converges.}

\section{The Newton-Raphson Pseudo-Demeaning Algorithm}
Now that we have derived an efficient way to update the structural parameters, this section is dedicated to present how the pseudo-demeaning approach can be  embedded into a standard Newton-Raphson routine. Remember that the Newton-Raphson routine requires to compute a gradient and Hessian in each iteration of the algorithm. Likewise (\ref{eq:betaupdate}) can be interpreted as a Newton-Raphson update based on a concentrated gradient and Hessian. Since those  are functions of the linear predictor $\boldsymbol{\eta}^{r}=   \mathbf{D} \boldsymbol{\alpha}^{r}     + \mathbf{X} \boldsymbol{\beta}^{r}$  we need to find an efficient way to update $\boldsymbol{\eta}^{r}$. The naive approach would be to recover an estimate of the fixed effects and use it to update the linear predictor. However this would be computationally inefficient.\footnote{For example one could apply a numerical solver for linear systems of equations as presented in section \ref{sec:Kaczmarz}.} We present a substantially less costly approach that  directly recovers the linear predictor. Therefore reconsider the reformulation of the Newton-Raphson update into the regression model
\begin{align}
	\label{eq:regression2}
\tilde{\boldsymbol{\nu}}^{r - 1} &= \widetilde{\mathbf{D}}^{r - 1} (\boldsymbol{\alpha}^{r} - \boldsymbol{\alpha}^{r - 1}) + \widetilde{\mathbf{X}}^{r - 1} (\boldsymbol{\beta}^{r}-\boldsymbol{\beta}^{r - 1}) \, .
\end{align}
The  normal equations of  system (\ref{eq:regression2}) are
\begin{align}
\label{eq:normalequations}
 \widetilde{\mathbf{X}}^{r - 1 \prime} \tilde{\boldsymbol{\nu}}^{r - 1} & = \widetilde{\mathbf{X}}^{r - 1 \prime} \widetilde{\mathbf{D}}^{r - 1} (\boldsymbol{\alpha}^{r} - \boldsymbol{\alpha}^{r - 1}) +  \widetilde{\mathbf{X}}^{r - 1 \prime} \widetilde{\mathbf{X}}^{r - 1} (\boldsymbol{\beta}^{r}-\boldsymbol{\beta}^{r - 1})	 \\
\widetilde{\mathbf{D}}^{r - 1 \prime}\tilde{\boldsymbol{\nu}}^{r - 1}  & =   	\widetilde{\mathbf{D}}^{r - 1 \prime} \widetilde{\mathbf{D}}^{r - 1} (\boldsymbol{\alpha}^{r} - \boldsymbol{\alpha}^{r - 1}) + \widetilde{\mathbf{D}}^{r - 1 \prime} \widetilde{\mathbf{X}}^{r - 1} (\boldsymbol{\beta}^{r}-\boldsymbol{\beta}^{r - 1})  \nonumber \, .
\end{align}
With some algebra on (\ref{eq:normalequations}) it can be shown that the residuals of the projected system (\ref{eq:projectedsystem}) are identical to the ones of the full system (\ref{eq:regression2}):
\begin{align}
\label{eq:residualequivalence}
\tilde{\boldsymbol{\nu}}^{r - 1} - \widetilde{\mathbf{X}}^{r - 1} (\boldsymbol{\beta}^{r}-\boldsymbol{\beta}^{r - 1}) - \widetilde{\mathbf{D}}^{r - 1} (\boldsymbol{\alpha}^{r}-\boldsymbol{\alpha}^{r - 1}) = \mathbf{M}_{\widetilde{\mathbf{D}}}^{r - 1} \tilde{\boldsymbol{\nu}}^{r - 1} - \mathbf{M}_{\widetilde{\mathbf{D}}}^{r - 1} \widetilde{\mathbf{X}}^{r - 1} (\boldsymbol{\beta}^r - \boldsymbol{\beta}^{r - 1}) \, .
\end{align}
Solving (\ref{eq:residualequivalence}) for $\boldsymbol{\eta}^{r}$ yields
\begin{align}
\label{eq:eta}
\boldsymbol{\eta}^{r} =    (\widetilde{\mathbf{W}}^{r-1})^{-1} \left(\tilde{\boldsymbol{\nu}}^{r - 1} - \mathbf{M}_{\widetilde{\mathbf{D}}}^{r - 1}  \tilde{\boldsymbol{\nu}}^{r - 1} - \mathbf{M}_{\widetilde{\mathbf{D}}}^{r - 1}  \tilde{\mathbf{X}}^{r - 1} (\boldsymbol{\beta}^r - \boldsymbol{\beta}^{r - 1}) \right) + \boldsymbol{\eta}^{r-1} \, .
\end{align}
Substituting the pseudo-demeaned variables for their approximations  $\ddot{\boldsymbol{\nu}}^{r}$ and $\ddot{\mathbf{X}}^{r}$ delivers  an efficient formula to obtain the linear predictor
\begin{align}
\label{eq:etaapproximation}
\boldsymbol{\eta}^{r} =   (\widetilde{\mathbf{W}}^{r-1})^{-1}  \left(\tilde{\boldsymbol{\nu}}^{r - 1} - \ddot{\boldsymbol{\nu}}^{r-1} - \ddot{\mathbf{X}}^{r-1} (\boldsymbol{\beta}^r - \boldsymbol{\beta}^{r - 1}) \right) + \boldsymbol{\eta}^{r-1} \, .
\end{align}
Bringing together all previously mentioned components the Newton-Raphson $k$-way pseudo-demeaning algorithm can be summarized by the following pseudo-code:

\begin{algorithm}
		\caption{Newton-Raphson with Pseudo-Demeaning}
\begin{algorithmic}[1]
	\State Initialize $\boldsymbol{\beta}^{0}$, $\boldsymbol{\eta}^{0}$, and  $r = 0$. 
	\Repeat
	\State Set $r = r + 1$.
	\State Compute the weights $\tilde{\mathbf{w}}^{r-1}$ and $\boldsymbol{\nu}^{r - 1}$ (see formula (\ref{eq:hessian})).
	\State Compute $\tilde{\boldsymbol{\nu}}^{r - 1}$ and $\widetilde{\mathbf{X}}^{r - 1}$ (see formula (\ref{eq:regression}).
	\State Compute $\ddot{\boldsymbol{\nu}}^{r - 1}$ and $\ddot{\mathbf{X}}^{r - 1}$ via AP using algorithm 1 or 2.
	\State Compute $(\boldsymbol{\beta}^{r} - \boldsymbol{\beta}^{r - 1}) = (\ddot{\mathbf{X}}^{r - 1 \prime}\ddot{\mathbf{X}}^{r - 1})^{-1}\ddot{\mathbf{X}}^{r - 1\prime}\ddot{\boldsymbol{\nu}}^{r - 1}$  (see formula (\ref{eq:betaupdate})) and update $\boldsymbol{\beta}^r$.
	\State Update $\boldsymbol{\eta}^{r}$ (see formula (\ref{eq:etaapproximation})).
	\Until{\textbf{convergence}.}
\end{algorithmic}
\end{algorithm}

Usually we are also interested in inference. Fortunately our algorithm is a maximum likelihood approach which facilitates the computation of different covariance estimators and allows for standard testing procedures. 
Let $r^{*}$ denote all quantities after convergences of algorithm 3. In order to compute the standard-errors of the structural parameters $\boldsymbol{\beta}^{r^*}$, we do not need the full Hessian  or full gradient.
The estimated variance-covariance matrix corresponding to the structural parameters $\boldsymbol{\beta}^{r^*}$ can be easily computed using the concentrated Hessian $\ddot{\mathbf{H}}$ after convergence:
\begin{equation}
	\hat{\mathbf{V}}_{emp}=\left(\ddot{\mathbf{X}}^{r^{*} \prime} \ddot{\mathbf{X}}^{r^{*}} \right)^{-1} = -\ddot{\mathbf{H}}^{-1} \, .
\end{equation}
A second estimator is based on the concentrated gradient $ \ddot{\mathbf{g}} = \ddot{\mathbf{X}}^{r^{*} \prime} \ddot{\boldsymbol{\nu}}^{r^{*}}$.
Therefore, define a $n \times p$ matrix $\ddot{\mathbf{G}} = [\ddot{\mathbf{g}}_1, \dots, \ddot{\mathbf{g}}_p]$, where the $n \times 1$  vector $\ddot{\mathbf{g}}_i = \ddot{\mathbf{X}}^{r^{*}}_i \odot \ddot{\boldsymbol{\nu}}^{r^{*}}$ contains the single contributions  of the $n$ observations to the $i$-th entry of concentrated gradient $\ddot{\mathbf{g}}$. 
The variance estimator becomes:
\begin{equation}
	\hat{\mathbf{V}}_{opg}=\left(\ddot{\mathbf{G}}' \ddot{\mathbf{G}}\right)^{-1} \, .
\end{equation} 
It is also known as the BHHH estimator or outer product of gradients estimator.
Thirdly, we present the sandwich estimator which is the standard estimator to obtain robust standard-errors
\begin{equation}
	\hat{\mathbf{V}}_{rob}=\ddot{\mathbf{H}}^{-1}\ddot{\mathbf{G}}'\ddot{\mathbf{G}}\ddot{\mathbf{H}}^{-1} \, .
\end{equation}

\section{Recovering the Fixed Effects Ex-Post}
\label{sec:Kaczmarz}
In some cases the researcher might not only require  estimates of the structural parameters but also of the fixed effects. System (\ref{eq:etaapproximation}) at convergence becomes
\begin{align}
\label{eq:etaconverged}
\boldsymbol{\eta}^{r^*} =    (\widetilde{\mathbf{W}}^{r^*-1})^{-1} \left(\tilde{\boldsymbol{\nu}}^{r^* - 1} - \ddot{\boldsymbol{\nu}}^{r^*-1} - \ddot{\mathbf{X}}^{r^*-1} (\boldsymbol{\beta}^{r^*}- \boldsymbol{\beta}^{r^* - 1}) \right) + \boldsymbol{\eta}^{r^*-1} \, .
\end{align}
Rearranging (\ref{eq:etaconverged}) yields a large and sparse system of linear equations
\begin{align}
\label{eq:systemalphaconvergedrearranged}
\mathbf{D} \boldsymbol{\alpha}^{r^*} = 
\underbrace{\boldsymbol{\eta}^{r^*}  -  \mathbf{X} \boldsymbol{\beta}^{r^*}}_{\mathbf{b}} \, ,
\end{align}
where $\mathbf{b}$ can be computed at low computational cost from already generated variables.
Since the analytical solution of (\ref{eq:systemalphaconvergedrearranged}) is inefficient and often infeasible, we propose  two numerical routines to solve the linear system of equations.\footnote{
	Note, that unlike to the analytical solution the numerical solvers do not require  $\mathbf{D}$ to have full rank. In order to get meaningful estimates for the fixed effects it is necessary to apply an estimable function to the solution \cite{Gaure2013theory}. However, for many ex-post analyses meaningful estimates are not required. }

The first solver we present is in spirit of the Gauss-Seidel algorithm  used by \citeN{Guimaraes2010}. We apply the same idea in order to compute the fixed effects by alternating between the normal equations corresponding to (\ref{eq:systemalphaconvergedrearranged}). Consider the case with three high-dimensional fixed effects $\boldsymbol{\alpha}_1, \boldsymbol{\alpha}_2$ and $\boldsymbol{\alpha}_3$.
The normal equations of system (\ref{eq:systemalphaconvergedrearranged}) are
\begin{align}
\label{eq:normalequation}
\mathbf{D}_1^{\prime} \mathbf{D}_1 \boldsymbol{\alpha}_1 + \mathbf{D}_1^{\prime} \mathbf{D}_2 \boldsymbol{\alpha}_2 + \mathbf{D}_1^{\prime} \mathbf{D}_3 \boldsymbol{\alpha}_3 & = \mathbf{D}_1^{\prime} \mathbf{b}\\
\mathbf{D}_2^{\prime} \mathbf{D}_1 \boldsymbol{\alpha}_1 + \mathbf{D}_2^{\prime} \mathbf{D}_2 \boldsymbol{\alpha}_2   + \mathbf{D}_2^{\prime} \mathbf{D}_3 \boldsymbol{\alpha}_3 & = \mathbf{D}_2^{\prime} \mathbf{b} \nonumber \\
\mathbf{D}_3^{\prime} \mathbf{D}_1 \boldsymbol{\alpha}_1  + \mathbf{D}_3^{\prime} \mathbf{D}_2 \boldsymbol{\alpha}_2 + \mathbf{D}_3^{\prime} \mathbf{D}_3 \boldsymbol{\alpha}_3  & = \mathbf{D}_3^{\prime} \mathbf{b} \nonumber
\end{align}
and can be rearranged to
\begin{align}
\label{eq:normalequationrearranged}
 \boldsymbol{\alpha}_1  & = (\mathbf{D}_1^{\prime} \mathbf{D}_1^{\prime})^{-1}\mathbf{D}_1^{\prime} (\mathbf{b} - \mathbf{D}_2 \boldsymbol{\alpha}_2 - \mathbf{D}_3 \boldsymbol{\alpha}_3)\\
 \boldsymbol{\alpha}_2  & = (\mathbf{D}_2^{\prime} \mathbf{D}_2^{\prime})^{-1}\mathbf{D}_2^{\prime} (\mathbf{b} - \mathbf{D}_1 \boldsymbol{\alpha}_1 - \mathbf{D}_3 \boldsymbol{\alpha}_3) \nonumber \\
  \boldsymbol{\alpha}_3  & = (\mathbf{D}_3^{\prime} \mathbf{D}_3^{\prime})^{-1}\mathbf{D}_3^{\prime} (\mathbf{b} - \mathbf{D}_1 \boldsymbol{\alpha}_1 - \mathbf{D}_2 \boldsymbol{\alpha}_2) \nonumber
\end{align}
The solver works as follows:  given some starting values for the fixed effects, we  alternate between the three normal equations. Fortunately, the single equations can be computed easily. The equation
$ \boldsymbol{\alpha}_i   = (\mathbf{D}_i^{\prime} \mathbf{D}_i^{\prime})^{-1}\mathbf{D}_i^{\prime} (\mathbf{b} - \mathbf{D}_{-i} \boldsymbol{\alpha}_{-i})$ is the group mean  of the vector $(\mathbf{b} - \mathbf{D}_{-i} \boldsymbol{\alpha}_{-i})$ by group $i$, where $\mathbf{D}_{-i} \boldsymbol{\alpha}_{-i}$ denotes all fixed effects contributions without the $i$-th .
Further, the part $\mathbf{D}_{-i} \boldsymbol{\alpha}_{-i}$ of the vector can be computed by stretching the corresponding fixed effects by their group identifiers. Algorithm 4 summarizes the procedure for an arbitrary number of fixed effects.

A second approach to solve the system (\ref{eq:systemalphaconvergedrearranged}) is the  Kaczmarz method \cite{Kaczmarz1937}. The Kaczmarz method belongs to the so called row-action methods and is suitable to solve large and
sparse systems (see  \citeN{Escalante2011alternating}). The idea is similar to the alternating projection methods described in section \ref{sec:pseudodemeaning}. Each equation of (\ref{eq:systemalphaconvergedrearranged}) defines a hyperplane and by alternating orthogonal projections on hyperplanes we can find the intersection. In our application the intersection are the fixed effects coefficients. 
Each projection of the $i$-th hyperplane onto the $i+1$-th hyperplane can be summarized as follows:
\begin{equation}	
	\boldsymbol{\rho}_{i+1}
	=
	\boldsymbol{\rho}_i
	+
	\frac{
		(\mathbf{b}_i- \langle\ \mathbf{d}_i, \boldsymbol{\rho}_i \rangle\ )}
	{||\mathbf{d}_i||_2^2}\mathbf{d}_i \, ,
\end{equation}
where $\boldsymbol{\rho}$ is a vector of length $l$, $\mathbf{d}_i$ and $\mathbf{b}_i$ denote the $i$-th row of $\mathbf{D}$ and $\mathbf{b}$ respectively, and $||\cdot||_2^2$ is the squared euclidean norm. Each row of $\mathbf{D}$  contains $K$ times the value one, such that the denominator can be simplified as follows
\begin{equation}
	\label{eq:kaczmarz}	
	\boldsymbol{\rho}_{i+1}
	=
	\boldsymbol{\rho}_i
	+
	\frac{
		(\mathbf{b}_i- \langle\ \mathbf{d}_i, \boldsymbol{\rho}_i \rangle\ )}
	{K}\mathbf{d}_i \, .
\end{equation}
Since $\mathbf{D}$ is sparse, the Kaczmarz updates can be computed at minimum memory. Algorithm 5 summarizes the procedure. In our applications we found that the first algorithm performs much faster.

\begin{algorithm}
	\caption{Alternating Between Normal Equations}
	\begin{algorithmic}[1]
		\State Set $j=1$, $\boldsymbol{\rho}_{j}= (\boldsymbol{\alpha}_{1j}, \dots, \boldsymbol{\alpha}_{Kj}) = \mathbf{0}_{K}$, $\boldsymbol{\rho}_{j - 1} = \boldsymbol{\rho}_{j} - \mathbf{1}_{K}$, and tolerance level $\epsilon$.
		\While{$||\boldsymbol{\rho}_{j} - \boldsymbol{\rho}_{j - 1}||_{2} \geq \epsilon$} 
		\For{$i=1, \ldots, K$}
		\State Compute $\boldsymbol{\alpha}_{ij}$ by computing the group mean over group $i$ of vector $(\mathbf{b}-\mathbf{D}_{-i}\boldsymbol{\alpha}_{-ij})$.
		\State Update $\boldsymbol{\rho}_{j}$ with new $\boldsymbol{\alpha}_{ij}$.
		\EndFor	
		\State Set $j = j + 1$.
		\EndWhile
		\State Set $\boldsymbol{\alpha}^{r^*} = \boldsymbol{\rho}_{j}$.
	\end{algorithmic}
\end{algorithm}

\begin{algorithm}
	\caption{Kaczmarz}
	\begin{algorithmic}[1]
		\State Set $j=1$, $\boldsymbol{\rho}_{j}= \mathbf{0}_{K}$, $\boldsymbol{\rho}_{j - 1} = \boldsymbol{\rho}_{j} - \mathbf{1}_{K}$, and tolerance level $\epsilon$.
		
		\While{$||\boldsymbol{\rho}_{j} - \boldsymbol{\rho}_{j - 1}||_{2} \geq \epsilon$} 
		\State Set $\boldsymbol{\rho}_{j0}=\boldsymbol{\rho}_{j}$.
		\For{$i=1, \ldots, n$}
		\State Compute $\boldsymbol{\rho}_{ji}$  (see formula \ref{eq:kaczmarz}).
		\EndFor	
		\State Set $j = j + 1$, $\boldsymbol{\rho}_{j} = \boldsymbol{\rho}_{jn}$.
		\EndWhile
		\State Set $\boldsymbol{\alpha}^{r^*} = \boldsymbol{\rho}_{j}$.
	\end{algorithmic}
\end{algorithm}

\section{Simulation}

To demonstrate the performance of our algorithm we consider two different simulation designs:
a two-way fixed effects logit model and a three-way (pseudo-) poisson model.\footnote{The application of a poisson estimator to a model with a continuous dependent model is popular in trade economics and called pseudo-poisson (see section \ref{sec:empirical} ).}  For both designs we analyse the exactness of the parameter estimates and the corresponding standard errors and  measure the computation times. We also consider different tolerance levels for the pseudo-demeaning algorithm.
All simulations were done with our  R-package \textit{alpaca} and a self-implementation of a GLM routine. The GLM routine is identical to our Newton-Raphson pseudo-demeaning algorithm except that we use dummy variables instead of alternating projections. This ensures the comparability between both algorithms. All computations were done on a Linux workstation (Ubuntu 16.04, Intel Xeon CPU 16 cores, 2.6 GHz, 64 GB RAM) using R version 3.4.4 \cite{R}.  For brevity we only report the results for the Neumann-Halperin algorithm.\footnote{We also performed simulations using Cimmino's approach and several acceleration schemes (\citeN{Hernandez2011}, \citeN{Gearhart1989}) that have been proposed in the literature. However, we did not find a superior algorithm. It is already well known that acceleration techniques can but do not necessarily accelerate (see e.g. \citeN{Hernandez2011}, \citeN{Escalante2011alternating}). Nevertheless, we observed that the classical Neumann-Halperin algorithm never performed worst.}

The first simulation experiment is a two-way fixed effects logit model where we generate data according to: 
\begin{equation}
\label{eq:dgplogit}
y_{it} = \mathbf{1}[\mathbf{x}_{it}^{\prime} \boldsymbol{\beta} + \alpha_{i} + \gamma_{t} + \epsilon_{it} > 0] \,, 
\end{equation}
where $i=1, \dots, N$, $t=1, \dots, T$, $x_{itp}$ is generated as iid. standard normal with $p=1, \dots, 3$, and $\epsilon_{it}$ is an iid. logistic error term with location zero and scale one,  $\alpha_{i} \sim \text{iid.}\; \mathcal{N}(\sum_{p=1}^{3} \bar{x}_{ip}, 1)$ and  $\gamma_{t}\sim \text{iid.}\; \mathcal{N}(\sum_{p=1}^{3} \bar{x}_{tp}, 1)$  and $\boldsymbol{\beta} = [1, - 1, 1]^{\prime}$. 

For the second experiment we consider a three-way fixed effects pseudo-poisson model. The data are generated according to
\begin{equation}
\label{eq:dgptrade}
Y_{ijt} = \exp(\alpha_{it} + \gamma_{jt} + \delta_{ij} + x_{ijt} \beta_1 + d_{ijt} \beta_2) \cdot \epsilon_{ijt} \,, 
\end{equation}
where $i=1, \dots, n$,  $j=1, \dots, n$, $t=1, \dots, T$, $x_{ijt}$ is generated as iid. standard normal, $d_{ijt} = 1 [\psi_{ijt} > 0]$, with $\psi_{ijt}$ is iid. standard normal, and $\epsilon_{ijt}$ is an iid. log-normal error-term with mean zero and variance one (on the log scale),  $\alpha_{it} \sim \text{iid.} \; \mathcal{N}( \bar{x}_{it}, 1)$,  $\gamma_{jt}\sim \text{iid.} \; \mathcal{N}(\bar{x}_{jt}, 1)$, $\delta_{ij}\sim \text{iid.} \; \mathcal{N}(\bar{x}_{ij}, 1)$,  and $\beta_1 = \beta_2 =1$. $\bar{x}_{.}$ denote the corresponding group means.

For both simulation experiments  we consider different combinations of $N$ and $T$  and generate $30$ different datasets for each combination.

\subsection{Exactness}
At first we investigate the exactness of the Newton-Raphson pseudo-demeaning algorithm because it is only an approximation. Therefore we measure how often the coefficients and standard errors differ from the exact dummy variable approach which gives us a reliable benchmark. We also consider different tolerance levels for the pseudo-demeaning algorithm. We only report the results for the first coefficient because the results of the other coefficients are similar. Tables 6.1 and 6.2 summarize the relative frequencies of  $\hat{\beta}_1$ and its standard error of the two-way logit model for different digits. The results of the PPML model can be found in the appendix \ref{sec:appendixppml} (Tables A.1 and A.2).
We observe that up to 5 digits the exact dummy approach and the  pseudo-demeaning deliver identical coefficients and standard errors irrespective of the chosen tolerance level. Additionally, we observe that the standard errors react more sensitive on the choose tolerance level. With respect to the exactness of the coefficients and the standard errors we recommend to use a tolerance level of $10^{-5}$.

\begin{table}[H]
		\label{tab:exactnesscoef}
	\begin{center}
		\caption{Exactness of  $\hat{\beta}_1$}
		\begin{tabular}{rrrrrrrrr}
			\toprule
			& N & T & $10^{-8}$ & $10^{-7}$ & $10^{-6}$ & $10^{-5}$  & $10^{-4}$ & $10^{-3}$ \\ 
			\midrule
5 digits	&	250 & 50 & 1.00 & 1.00 & 1.00 & 1.00 & 1.00 & 1.00 \\ 
	&	250 & 100 & 1.00 & 1.00 & 1.00 & 1.00 & 1.00 & 1.00 \\ 
	&	500 & 50 & 1.00 & 1.00 & 1.00 & 1.00 & 1.00 & 1.00 \\ 
	&	500 & 100 & 1.00 & 1.00 & 1.00 & 1.00 & 1.00 & 1.00 \\ 
	&	500 & 250 & 1.00 & 1.00 & 1.00 & 1.00 & 1.00 & 1.00 \\ 
	\midrule
8 digits	&	250 & 50 & 1.00 & 1.00 & 1.00 & 1.00 & 1.00 & 0.97 \\ 
	&	250 & 100 & 1.00 & 1.00 & 1.00 & 1.00 & 1.00 & 1.00 \\ 
	&	500 & 50 & 1.00 & 1.00 & 1.00 & 1.00 & 1.00 & 0.93 \\ 
	&	500 & 100 & 1.00 & 1.00 & 1.00 & 1.00 & 1.00 & 1.00 \\ 
	&	500 & 250 & 1.00 & 1.00 & 1.00 & 1.00 & 1.00 & 0.97 \\ 
	\midrule
16 digits	&	250 & 50 & 0.53 & 0.10 & 0.00 & 0.00 & 0.00 & 0.00 \\ 
	&	250 & 100 & 0.40 & 0.27 & 0.03 & 0.00 & 0.00 & 0.00 \\ 
	&	500 & 50 & 0.60 & 0.20 & 0.00 & 0.00 & 0.00 & 0.00 \\ 
	&	500 & 100 & 0.47 & 0.57 & 0.00 & 0.00 & 0.00 & 0.00 \\ 
	&	500 & 250 & 0.60 & 0.50 & 0.13 & 0.00 & 0.00 & 0.00 \\ 
		\bottomrule
\end{tabular}
\end{center}
\footnotesize{\emph{Note}: Two-way logit model; measurement of exactness frequencies relative to dummy variable approach  up to $5$, $8$ and $16$ digits over $30$ datasets per $N-T$ combination; used Neumann-Halperin projection with different tolerance levels.}
\end{table}

\begin{table}[H]
	\label{tab:exactnessse}
	\begin{center}
		\caption{Exactness of  $se(\hat{\beta}_1)$}
		\begin{tabular}{rrrrrrrrr}
			\toprule
			& N & T & $10^{-8}$ & $10^{-7}$ & $10^{-6}$ & $10^{-5}$  & $10^{-4}$ & $10^{-3}$ \\ 
			\midrule
5 digits	&	250 & 50 & 1.00 & 1.00 & 1.00 & 1.00 & 1.00 & 1.00 \\ 
	&	250 & 100 & 1.00 & 1.00 & 1.00 & 1.00 & 1.00 & 1.00 \\ 
	&	500 & 50 & 1.00 & 1.00 & 1.00 & 1.00 & 1.00 & 1.00 \\ 
	&	500 & 100 & 1.00 & 1.00 & 1.00 & 1.00 & 1.00 & 1.00 \\ 
	&	500 & 250 & 1.00 & 1.00 & 1.00 & 1.00 & 1.00 & 1.00 \\ 
	\midrule
8 digits	&	250 & 50 & 1.00 & 1.00 & 1.00 & 1.00 & 0.60 & 0.03 \\ 
	&	250 & 100 & 1.00 & 1.00 & 0.93 & 0.93 & 0.30 & 0.07 \\ 
	&	500 & 50 & 1.00 & 1.00 & 1.00 & 1.00 & 0.70 & 0.00 \\ 
	&	500 & 100 & 1.00 & 1.00 & 1.00 & 1.00 & 0.40 & 0.00 \\ 
	&	500 & 250 & 1.00 & 1.00 & 1.00 & 1.00 & 0.17 & 0.00 \\ 
	\midrule
16 digits	&	250 & 50 & 0.00 & 0.00 & 0.00 & 0.00 & 0.00 & 0.00 \\ 
	&	250 & 100 & 0.00 & 0.00 & 0.00 & 0.00 & 0.00 & 0.00 \\ 
	&	500 & 50 & 0.00 & 0.00 & 0.00 & 0.00 & 0.00 & 0.00 \\ 
	&	500 & 100 & 0.00 & 0.00 & 0.00 & 0.00 & 0.00 & 0.00 \\ 
	&	500 & 250 & 0.00 & 0.00 & 0.00 & 0.00 & 0.00 & 0.00 \\ 
	\bottomrule
\end{tabular}
\end{center}
\footnotesize{\emph{Note}: Two-way logit model; measurement of exactness frequencies relative to dummy variable approach  up to $5$, $8$ and $16$ digits over $30$ datasets per $N-T$ combination; used Neumann-Halperin projection with different tolerance levels.}
\end{table}

\subsection{Computation Times}

Next we consider the computation times of the naive dummy variable approach and the Newton-Raphson pseudo-demeaning algorithm for different tolerance levels. Table 6.3 shows the dramatic increase of the computation time of the dummy variable approach. Whereas the dummy variable approach takes roughly $5$ minutes to estimate a two-way fixed effects logit model with $125,000$ observations and $750$ fixed effects, our approach requires less than $1$ second. For higher combinations we only report the computation times obtained by the Newton-Raphson pseudo-demeaning algorithm. Even in the largest dataset consisting of $10$ million observations and including $11,000$ fixed effects our routine is able to estimate the model in roughly $1.5$ minutes for the tightest tolerance level of $10^{-8}$ and only $1$ minute for the loosest. The computation times of the three-way PPML model  can be found in the appendix \ref{sec:appendixppml} (Table A.3).\footnote{Note that it is also possible to parallelize the pseudo-demeaning algorithm  over the number of model variables to gain a further speed advantage.}

\begin{table}[H]
	\label{tab:timeglmalpacalogit}
	\begin{center}
\caption{Average Computation Time in Seconds}
	\begin{tabular}{rrrrrrrrr}
		\toprule
		                      &         &        &                      \multicolumn{6}{c}{Alpaca}                       \\
		\cmidrule(lr){4-9}  N &       T &  Dummy & $10^{-8}$ & $10^{-7}$ & $10^{-6}$ & $10^{-5}$ & $10^{-4}$ & $10^{-3}$ \\ \midrule
		               250 &   50 &   5.06 &      0.09 &      0.10 &      0.11 &      0.10 &      0.09 &      0.09 \\
		               250 &  100 &  14.03 &      0.22 &      0.20 &      0.16 &      0.15 &      0.18 &      0.15 \\
		               500 &   50 &  33.58 &      0.21 &      0.20 &      0.19 &      0.19 &      0.14 &      0.14 \\
		               500 &  100 &  79.75 &      0.39 &      0.38 &      0.35 &      0.30 &      0.32 &      0.32 \\
		               500 &  250 & 302.04 &      0.95 &      0.89 &      0.81 &      0.80 &      0.74 &      0.73 \\
		              1000 &   50 &    -    &      0.36 &      0.37 &      0.35 &      0.33 &      0.31 &      0.30 \\
		              1000 &  100 &    -    &      0.67 &      0.64 &      0.61 &      0.60 &      0.57 &      0.54 \\
		              1000 &  250 &    -    &      1.81 &      1.70 &      1.58 &      1.55 &      1.46 &      1.47 \\
		              1000 &  500 &    -    &      3.81 &      3.85 &      3.51 &      3.37 &      3.08 &      2.77 \\
		              5000 &   50 &    -    &      1.88 &      1.82 &      1.81 &      1.67 &      1.52 &      1.45 \\
		              5000 &  100 &    -    &      3.92 &      3.67 &      3.42 &      3.46 &      3.20 &      2.81 \\
		              5000 &  250 &    -    &      9.85 &      9.56 &      8.60 &      8.21 &      7.61 &      7.05 \\
		              5000 &  500 &     -   &     20.14 &     19.76 &     18.04 &     17.43 &     15.75 &     15.24 \\
		              5000 & 1000 &     -   &     40.57 &     39.55 &     36.46 &     34.93 &     32.37 &     30.28 \\
		             10000 &   50 &     -   &      3.92 &      3.78 &      3.60 &      3.40 &      3.18 &      3.02 \\
		             10000 &  100 &    -    &      8.13 &      7.63 &      7.24 &      6.87 &      6.34 &      5.85 \\
		             10000 &  250 &     -   &     20.42 &     19.67 &     17.79 &     17.25 &     15.94 &     14.65 \\
		             10000 &  500 &    -    &     40.83 &     39.68 &     36.56 &     35.32 &     32.68 &     30.38 \\
		             10000 & 1000 &    -    &     87.32 &     85.88 &     78.74 &     75.75 &     70.31 &     65.59 \\ \bottomrule
	\end{tabular}
	\end{center}
\footnotesize{\emph{Note}: Two-way logit model; average computation times in seconds over $30$ datasets per $N-T$ combination; \textit{Dummy} refers to maximum likelihood estimation with dummy variables to account for the unobserved heterogeneity; \textit{Alpaca}  refers to the Newton-Raphson pseudo-demeaning algorithm (used Neumann-Halperin projection with different tolerance levels).}
\end{table}

\section{Empirical Example}
\label{sec:empirical}
The workhorse approach to estimate structural gravity models is  the pseudo-poisson maximum likelihood estimator proposed by \citeN{Silva2006}.\footnote{The estimator is a poisson maximum likelihood estimator applied to a non-poisson distributed dependent variable (trade flows are positive and continuous). See \citeN{Gourieroux1984} why this application is valid.}
A standard panel data gravity model takes the following form
\begin{equation*}
Y_{ijt} = \exp(\alpha_{it} + \alpha_{jt} + \alpha_{ij} + \mathbf{x}_{ijt}^{\prime} \boldsymbol{\beta}) \epsilon_{ijt}
\end{equation*}
where $Y_{ijt}$ denotes the trade flows from exporter $i$ to importer $j$ at time $t$, $\alpha_{it}$ is an exporter-time fixed effect, $\alpha_{jt}$ is an importer-time fixed effect and $\alpha_{ij}$ is an exporter-importer (dyadic) fixed effect, $\mathbf{x}_{ijt}$ is a vector of further regressors and $\boldsymbol{\beta}$ the corresponding parameter vector.
For model specifications where the three fixed effects are high-dimensional, researchers may face computational limits.  For example \citeN{Glick2016} used a panel dataset with over 200 countries trading for 65 years. In this case, the dataset consists of $879,794$ observations (after dropping all zero trade flows) and a three-way fixed effects specification would require roughly $11,000$ exporter-time and importer-time fixed effects, respectively, as well as roughly $34,000$ dyadic fixed effects. Due to the lack of a feasible software routine \citeN{Glick2016} estimated a three-way fixed effects log-linear specification instead of the desired PPML counterpart. 

Recently, \citeN{Larch2017} proposed a feasible PPML algorithm based on  \citeN{Guimaraes2010} that can handle high-dimensional three-way fixed effects. With this tool at hand, they are able to estimate the model of \citeN{Glick2016} with the full set of fixed effects. Unlike the log linear specification, PPML is able to deal with zero trade flows, thus it makes sense to use the entire data set. Following \citeN{Larch2017} we replace all missing trade flows with zeros resulting in a full data set of roughly 3 million observations. We replicate table A2 to show that in the context of PPML our Newton-Raphson pseudo-demeaning algorithm is an alternative to the Gauss-Seidel IRLS routine of \citeN{Larch2017}.

\citeN{Glick2016} estimate the following theory-consistent gravity model
\begin{equation*}
X_{ijt}= \exp(\gamma CU_{ijt} + Z_{ijt}^{\prime}\boldsymbol{\beta}  + \lambda_{it} + \psi_{jt} + \delta_{ij}) u_{ijt} \, ,
\end{equation*}
where $X_{ijt}$ denotes the nominal value of bilateral exports from exporter $i$ to importer $j$ at year $t$, $CU_{ijt}$ is dummy variable, specifying whether $i$ and $j$ use the same currency at time $t$, $Z_{ijt}$ are further control variables, $\lambda_{it}$ denotes a time-varying exporter fixed effect, $\psi_{jt}$ a time-varying importer fixed effect, and $\delta_{ij}$ is a dyadic fixed effect.
Table 7.1 reproduces table A2 from \citeN{Larch2017}  using our Newton-Raphson pseudo-demeaning algorithm.\footnote{For a detailed discussion of the estimation results please consider \citeN{Larch2017}.}
Depending on the model specification, we are able to estimate the model in $40$ up to $110$ seconds. 

Further we want to show that our routine is also able to deal with many regressors. Therefor we replicate table 6 of \citeN{Glick2016} where they test the symmetry assumption between entries and exits of countries from a currency union using joint hypotheses. Again we use the full data set and the following specification
\begin{equation*}
X_{ijt}= \exp\left(\sum_{k = - 14}^{14} \theta_{k} \text{CUENTRY}_{ij(t-k)} + \sum_{k = - 14}^{14} \phi_{k} \text{CUEXIT}_{ij(t-k)} + Z_{ijt}^{\prime}\boldsymbol{\beta}  + \lambda_{it} + \psi_{jt} + \delta_{ij}\right) u_{ijt} \, ,
\end{equation*}
where $\text{CUEXIT}_{ij(t - k)}$ is one if country $i$ and $j$ entered a currency union at time $t - k$ and $\text{CUEXIT}_{ij(t - k)}$ is one if country $i$ and $j$ exited a currency union at time $t - k$.

Table 7.2 summarizes the results of seven different Wald-tests. The tests require to estimate  unrestricted models with $60$ and $89$ regressors.  Contrary to \citeN{Glick2016} we reject the null of symmetry in most of the cases. To sum up, our routine is able to estimate models with many high-dimensional fixed effects in a reasonable amount of time even in the presence of many regressors.

 \begin{landscape}
 	\begin{table}
 		\label{tab:timetrade3way}
 		\caption{Empirical Results}
 		\begin{center}
 			\renewcommand{\arraystretch}{0.8}
 			\begin{tabular}{llrrr}
 				\toprule
 				\cmidrule(lr){3-5}                     &      &                  All CUs &              Disagg. EMU &              Disagg. CUs \\ \midrule
 				\textbf{All Currency Unions}           & coef &                   0.1531 &                          &  \\
 				                                       & se   & (0.0102, 0.0432, 0.0810) &                          &  \\ \midrule
 				\textbf{All Non-EMU Currency Unions}   & coef &                          &                   0.7276 &  \\
 				                                       & se   &                          & (0.0255, 0.1103, 0.1786) &  \\ \midrule
 				\textbf{EMU }                          & coef &                          &                   0.0521 &                   0.0489 \\
 				                                       & se   &                          & (0.0103, 0.0427, 0.0947) & (0.0103, 0.0423, 0.0946) \\ \midrule
 				\textbf{CFA Franc Zone}                & coef &                          &                          &                    -0.1256 \\
 				                                       & se   &                          &                          & (0.0997, 0.3373, 0.3551) \\ \midrule
 				\textbf{East Caribbean Currency Union} & coef &                          &                          &                  -0.8773 \\
 				                                       & se   &                          &                          & (0.0835, 0.2961, 0.2960) \\ \midrule
 				\textbf{Aussie}                        & coef &                          &                          &                   0.3845 \\
 				                                       & se   &                          &                          & (0.1188, 0.2427, 0.2186) \\ \midrule
 				\textbf{British} $\textsterling$       & coef &                          &                          &                   1.0600 \\
 				                                       & se   &                          &                          & (0.0347, 0.1455, 0.2355) \\ \midrule
 				\textbf{French Franc}                  & coef &                          &                          &                   2.0957 \\
 				                                       & se   &                          &                          & (0.0630, 0.2290, 0.3093) \\ \midrule
 				\textbf{Indian Ruppee}                 & coef &                          &                          &                   0.1697 \\
 				                                       & se   &                          &                          & (0.1470, 0.3713, 0.2660) \\ \midrule
 				\textbf{US} $\$$                       & coef &                          &                          &                    0.0183 \\
 				                                       & se   &                          &                          & (0.0215, 0.0655, 0.0509) \\ \midrule
 				\textbf{Other CUs}                     & coef &                          &                          &                    0.7660 \\
 				                                       & se   &                          &                          & (0.0533, 0.1848, 0.2474) \\ \midrule
 				Importer-time fixed effects            &      &                 $11,277$ &                 $11,277$ &                 $11,277$ \\
 				Exporter-time fixed effects            &      &                 $11,227$ &                 $11,227$ &                 $11,227$ \\
 				Dyadic fixed effects                   &      &                 $34,104$ &                 $34,104$ &                 $34,104$ \\ \midrule
 				time (in sec.)                         &      &                       41 &                       49 &                       107 \\
 				iterations                             &      &                        9 &                        9 &                       10 \\ \bottomrule
 			\end{tabular}
 		\end{center}
 		\footnotesize{\emph{Note}: After dropping observations that do not contribute to the log-likelihood, we end up with roughly 1.6 million observations; two further control variables: regional FTA membership and current colony/colonizer; robust (sandwich estimator), clustered standard errors by country pairs, and multi-way clustered standard errors by importer, exporter, and time in parenthesis; Neumann-Halperin projection with tolerance level $10^{-5}$.}
 	\end{table}
\end{landscape}

\begin{table}
	\label{tab:symmetry}
	\caption{Entry -- Exit Symmetry}
	\begin{center}
	\begin{tabular}{lr}
		\toprule
		Hypothesis                                  & Wald-test \\ \midrule
		After any CU Entry = - After any CU Exit?   &  27.5 (0.0165)\\
		Before any CU Entry = - Before any CU Exit? &  16.8 (0.2665)\\
		Both                                        &  149.8 (0.0000)\\
		\textit{Number of regressors: 60, Time: 10 minutes}&\\
		\midrule
		After non-EMU CU Entry = After EMU Entry?   &  29.0 (0.0106)\\
		Before non-EMU CU Entry = Before EMU Entry? &  52.5 (0.0000)\\
		Both                                        &  80.3 (0.0000)\\
		After non-EMU CU Exit = - After EMU Entry?  &  28.3 (0.0132)\\
		\textit{Number of regressors: 89, Time: 15 minutes}&\\ \bottomrule
	\end{tabular}
	\end{center}
\footnotesize{\emph{Note}: Wald-tests based on robust covariances, reported test-statistics and p-values in parenthesis; two further control variables: regional FTA membership and current colony/colonizer; Neumann-Halperin projection with tolerance level $10^{-5}$.}
\end{table}

\section{Conclusion}
\normalsize
We present a new algorithm for the maximum likelihood estimation of generalized linear models with a high-dimensional $k$-way error component. Our approach is straightforward since it resembles the classical ``within'' transformation used in linear regression models.
To be more precise the algorithm incorporates a special pseudo-demeaning procedure into a standard Newton-Raphson estimation routine such that the updates of the structural parameters are separated from the high-dimensional fixed effects. We show that our algorithm delivers almost identical estimates compared to the classical maximum likelihood estimation routine with dummy variables. Whereas the latter quickly becomes either time demanding or even infeasible, our algorithm  is memory efficient and thus offers new possibilities and  reliefs to researchers. It even allows to estimate models with many observations and high-dimensional fixed effects on a standard computer. To make our algorithm available for empirical research we provide an R-package \textit{alpaca}. 

In our empirical example we estimate a structural gravity model with three high-dimensional fixed effects  to demonstrate one possible area of application for our algorithm.

Although this paper focuses on generalized linear models the proposed procedure might be adjustable to other non-linear models whenever the Newton-Raphson update can be represented as a weighted least squares problem.

Our algorithm also encourages the development of bias-correction methods for $k$-way error component models since it alleviates the computational burden of extensive Monte-Carlo experiments.  For two-way fixed effects panel data models it is already straightforward to combine our algorithm with the  bias correction proposed by \citeN{Fernandez2016} to reduce the well known incidental parameter bias.

\newpage

\begin{LARGE}
	{\textbf{Appendix}}
\end{LARGE}

\appendix
\numberwithin{table}{section}

\section{Simulation Results of the Three-Way PPML Model}
\label{sec:appendixppml}

\begin{table}[H]
	\label{tab:exactnesscoefppml}
	\begin{center}
		\caption{Exactness of  $\hat{\beta}_1$}
		\begin{tabular}{rrrrrrrrr}
			\toprule
			& N & T & $10^{-8}$ & $10^{-7}$ & $10^{-6}$ & $10^{-5}$  & $10^{-4}$ & $10^{-3}$ \\ 
			\midrule
			5 digits	&	10 & 5 & 1.00 & 1.00 & 1.00 & 1.00 & 1.00 & 1.00 \\ 
			&	10 & 10 & 1.00 & 1.00 & 1.00 & 1.00 & 1.00 & 1.00 \\ 
			&	10 & 25 & 1.00 & 1.00 & 1.00 & 1.00 & 1.00 & 1.00 \\ 
			&	10 & 50 & 1.00 & 1.00 & 1.00 & 1.00 & 1.00 & 1.00 \\ 
			&	25 & 5 & 1.00 & 1.00 & 1.00 & 1.00 & 1.00 & 1.00 \\ 
			&	25 & 10 & 1.00 & 1.00 & 1.00 & 1.00 & 1.00 & 1.00 \\ 
			&	25 & 25 & 1.00 & 1.00 & 1.00 & 1.00 & 1.00 & 1.00 \\ 
			&	25 & 50 & 1.00 & 1.00 & 1.00 & 1.00 & 1.00 & 1.00 \\ 
			\midrule
			8 digits	&	10 & 5 & 1.00 & 1.00 & 1.00 & 1.00 & 1.00 & 1.00 \\ 
			&	10 & 10 & 1.00 & 1.00 & 1.00 & 1.00 & 1.00 & 1.00 \\ 
			&	10 & 25 & 1.00 & 1.00 & 1.00 & 1.00 & 1.00 & 1.00 \\ 
			&	10 & 50 & 1.00 & 1.00 & 1.00 & 1.00 & 1.00 & 1.00 \\ 
			&	25 & 5 & 1.00 & 1.00 & 1.00 & 1.00 & 1.00 & 1.00 \\ 
			&	25 & 10 & 1.00 & 1.00 & 1.00 & 1.00 & 1.00 & 1.00 \\ 
			&	25 & 25  & 1.00 & 1.00 & 1.00 & 1.00 & 1.00 & 1.00 \\ 
			&	25 & 50  & 1.00 & 1.00 & 1.00 & 1.00 & 1.00 & 1.00 \\ 
			\midrule
			16 digits	&	10 & 5 & 0.00 & 0.00 & 0.00 & 0.00 & 0.00 & 0.00 \\ 
			&	10 & 10 & 0.03 & 0.03 & 0.03 & 0.03 & 0.03 & 0.03 \\ 
			&	10  & 25 & 0.00 & 0.00 & 0.00 & 0.00 & 0.00 & 0.00 \\ 
			&	10 & 50 & 0.00 & 0.00 & 0.00 & 0.00 & 0.00 & 0.00 \\ 
			&	25 & 5 & 0.00 & 0.00 & 0.00 & 0.00 & 0.00 & 0.00 \\ 
			&	25 & 10 & 0.03 & 0.03 & 0.03 & 0.03 & 0.03 & 0.03 \\ 
			&	25 & 25 & 0.00 & 0.00 & 0.00 & 0.00 & 0.00 & 0.00 \\ 
			&	25 & 50 & 0.00 & 0.00 & 0.00 & 0.00 & 0.00 & 0.00 \\ 
			\bottomrule
		\end{tabular}
	\end{center}
	\footnotesize{\emph{Note}: Three-way PPML model; measurement of exactness frequencies relative to dummy variable approach  up to $5$, $8$ and $16$ digits over $30$ datasets per $N-T$ combination; used Neumann-Halperin projection with different tolerance levels.}
\end{table}

\newpage
\begin{table}[H]
	\label{tab:exactnessseppml}
	\begin{center}
		\caption{Exactness of $se(\hat{\beta}_1)$}
		\begin{tabular}{rrrrrrrrr}
			\toprule
			& N & T & $10^{-8}$ & $10^{-7}$ & $10^{-6}$ & $10^{-5}$  & $10^{-4}$ & $10^{-3}$ \\ 
			\midrule
			5 digits	&	10 & 5 & 1.00 & 1.00 & 1.00 & 1.00 & 1.00 & 1.00 \\ 
			&	10 & 10 & 1.00 & 1.00 & 1.00 & 1.00 & 1.00 & 1.00 \\ 
			&	10 & 25 & 1.00 & 1.00 & 1.00 & 1.00 & 1.00 & 1.00 \\ 
			&	10 & 50 & 1.00 & 1.00 & 1.00 & 1.00 & 1.00 & 1.00 \\ 
			&	25 & 5 & 1.00 & 1.00 & 1.00 & 1.00 & 1.00 & 1.00 \\ 
			&	25 & 10 & 1.00 & 1.00 & 1.00 & 1.00 & 1.00 & 1.00 \\ 
			&	25 & 25 & 1.00 & 1.00 & 1.00 & 1.00 & 1.00 & 1.00 \\ 
			&	25 & 50 & 1.00 & 1.00 & 1.00 & 1.00 & 1.00 & 1.00 \\
			\midrule 
			8 digits	&	10 & 5 & 0.97 & 0.97 & 0.97 & 0.97 & 0.97 & 0.97 \\ 
			&	10 & 10 & 1.00 & 1.00 & 1.00 & 1.00 & 1.00 & 1.00 \\ 
			&	10 & 25 & 1.00 & 1.00 & 1.00 & 1.00 & 1.00 & 1.00 \\ 
			&	10 & 50 & 1.00 & 1.00 & 1.00 & 1.00 & 1.00 & 1.00 \\ 
			&	25 & 5 & 1.00 & 1.00 & 1.00 & 1.00 & 1.00 & 1.00 \\ 
			&	25 & 10 & 1.00 & 1.00 & 1.00 & 1.00 & 1.00 & 1.00 \\ 
			&	25 & 25 & 1.00 & 1.00 & 1.00 & 1.00 & 1.00 & 1.00 \\ 
			&	25 & 50 & 1.00 & 1.00 & 1.00 & 1.00 & 1.00 & 1.00 \\ 
			\midrule
			16 digits	&	10 & 50 & 0.00 & 0.00 & 0.00 & 0.00 & 0.00 & 0.00 \\ 
			&	10 & 10 & 0.00 & 0.00 & 0.00 & 0.00 & 0.00 & 0.00 \\ 
			&	10 & 25 & 0.00 & 0.00 & 0.00 & 0.00 & 0.00 & 0.00 \\ 
			&	10 & 50 & 0.00 & 0.00 & 0.00 & 0.00 & 0.00 & 0.00 \\ 
			&	25 & 5 & 0.00 & 0.00 & 0.00 & 0.00 & 0.00 & 0.00 \\ 
			&	25 & 10 & 0.00 & 0.00 & 0.00 & 0.00 & 0.00 & 0.00 \\ 
			&	25 & 25 & 0.00 & 0.00 & 0.00 & 0.00 & 0.00 & 0.00 \\ 
			&	25 & 50 & 0.00 & 0.00 & 0.00 & 0.00 & 0.00 & 0.00 \\ 
			\bottomrule
		\end{tabular}
	\end{center}
	\footnotesize{\emph{Note}: Three-way PPML model; measurement of exactness frequencies relative to dummy variable approach  up to $5$, $8$ and $16$ digits over $30$ datasets per $N-T$ combination; used Neumann-Halperin projection with different tolerance levels.}
\end{table}

\begin{table}[H]
	\label{tab:timeglmalpacappml}
	\begin{center}
		\caption{Average Computation Time in Seconds}
		\begin{tabular}{rrrrrrrrr}
			\toprule
			&         &        &                      \multicolumn{6}{c}{Alpaca}                       \\
			\cmidrule(lr){4-9}  N &       T &  Dummy & $10^{-8}$ & $10^{-7}$ & $10^{-6}$ & $10^{-5}$ & $10^{-4}$ & $10^{-3}$ \\ \midrule
			10 & 5 & 0.19 & 0.07 & 0.06 & 0.06 & 0.07 & 0.06 & 0.05 \\ 
			10 & 10 & 0.66 & 0.06 & 0.06 & 0.06 & 0.07 & 0.07 & 0.07 \\ 
			10 & 25 & 4.93 & 0.14 & 0.13 & 0.14 & 0.14 & 0.15 & 0.14 \\ 
			10 & 50 & 35.82 & 0.23 & 0.24 & 0.25 & 0.25 & 0.25 & 0.26 \\ 
			25 & 5 & 14.49 & 0.17 & 0.18 & 0.16 & 0.17 & 0.16 & 0.15 \\ 
			25 & 10 & 46.44 & 0.26 & 0.27 & 0.28 & 0.27 & 0.26 & 0.25 \\ 
			25 & 25 & 320.65 & 0.51 & 0.52 & 0.51 & 0.50 & 0.53 & 0.51 \\ 
			25 & 50 & 2047.95 & 1.07 & 1.08 & 1.06 & 1.07 & 1.07 & 1.07 \\ 
			50 & 5 & - & 0.48 & 0.47 & 0.50 & 0.48 & 0.47 & 0.49 \\ 
			50 & 10 & - & 0.98 & 0.97 & 0.96 & 0.98 & 0.99 & 0.97 \\ 
			50 & 25 &  - & 2.01 & 2.07 & 2.10 & 2.03 & 2.13 & 2.04 \\ 
			50 & 50 & - & 4.37 & 4.40 & 4.31 & 4.69 & 4.57 & 4.56 \\ 
			100 & 5 & - & 2.16 & 2.20 & 2.20 & 2.06 & 2.18 & 2.04 \\ 
			100 & 10 & - & 3.92 & 3.63 & 3.60 & 3.57 & 3.61 & 3.64 \\ 
			100 & 25 & - & 11.25 & 11.01 & 11.29 & 11.37 & 10.92 & 10.95 \\ 
			100 & 50 & - & 28.53 & 28.57 & 28.57 & 28.49 & 28.53 & 28.27 \\ 
			200 & 5 &  - & 8.53 & 8.51 & 8.56 & 8.55 & 8.59 & 8.48 \\ 
			200 & 10 & - & 26.62 & 26.88 & 26.51 & 26.66 & 26.35 & 26.39 \\ 
			200 & 25 & - & 65.48 & 65.59 & 65.78 & 66.63 & 65.02 & 64.91 \\ 
			200 & 50 & -  & 140.91 & 139.52 & 141.60 & 140.78 & 140.22 & 140.33 \\ 
			\bottomrule
		\end{tabular}
	\end{center}
	\footnotesize{\emph{Note}: Three-way PPML model; average computation times in seconds over $30$ datasets per $N-T$ combination; \textit{Dummy} refers to maximum likelihood estimation with dummy variables to account for the unobserved heterogeneity; \textit{Alpaca}  refers to the Newton-Raphson pseudo-demeaning algorithm (used Neumann-Halperin projection with different tolerance levels).}
\end{table}

\clearpage
\ifx\undefined\bysame
\newcommand{\bysame}{\leavevmode\hbox to\leftmargin{\hrulefill\,\,}}
\fi

\end{document}